\documentclass[12pt]{article}
\usepackage[dvips]{graphics}

\def\beq   {\begin{equation}}
\def\eeq   {\end{equation}}
\def\beqd  {\begin{displaymath}}
\def\eeqd  {\end{displaymath}}
\def\beqaa {\begin{eqnarray}}
\def\eeqaa {\end{eqnarray}}

\def\noi {\noindent}

\def\ti  {\tilde}

\def\sq  {\ti q}
\def\st  {\ti t}
\def\sb  {\ti b}
\def\sg  {\ti g}

\def\nt  {\tilde\chi^0}
\def\ch  {\tilde\chi^\pm}
\def\chp {\tilde\chi^+}

\def\a   {\alpha}
\def\b   {\beta}
\def\t   {\theta}

\def\sz{\ifmmode{\tilde{\chi}^0} \else{$\tilde{\chi}^0$} \fi}
\def\sw{\ifmmode{\tilde{\chi}} \else{$\tilde{\chi}$} \fi}

\newcommand{\gsim}{\;\raisebox{-0.9ex}
           {$\textstyle\stackrel{\textstyle >}{\sim}$}\;}
\newcommand{\lsim}{\;\raisebox{-0.9ex}{$\textstyle\stackrel{\textstyle<}
           {\sim}$}\;}

\begin{document}
\pagestyle{plain}

\vspace*{-1cm} 
\begin{flushright}
  TGU-34 \\
  UWThPh-2004-26 \\
  hep-ph/0409347
\end{flushright}

\vspace*{1.4cm}

\begin{center}

{\Large {\bf
Impact of CP phases on the search for top and bottom squarks 
}}\footnote
{Invited talk presented by K. Hidaka in The 12th International 
Conference on Supersymmetry and Unification of Fundamental Interactions 
(SUSY04), June 17-23, 2004, Tsukuba, Japan.}\\

\vspace{10mm}

{\large 
A.~Bartl$^a$, S.~Hesselbach$^a$, {\underline {K.~Hidaka}}$^b$, 
T.~Kernreiter$^a$ and W.~Porod$^c$}

\vspace{6mm}

\begin{tabular}{l}
$^a${\it Institut f\"ur Theoretische Physik, Universit\"at Wien, A-1090
Vienna, Austria}\\
$^b${\it Department of Physics, Tokyo Gakugei University, Koganei,
Tokyo 184--8501, Japan}\\
$^c${\it Institut f\"ur Theoretische Physik, Universit\"at Z\"urich, 
CH-8057 Z\"urich, Switzerland}
\end{tabular}

\end{center}

\vfill

\begin{abstract} 
We study the decays of top squarks ($\tilde t_{1,2}$)
and bottom squarks ($\tilde b_{1,2}$) in the Minimal
Supersymmetric Standard Model (MSSM) with complex
parameters $A_t,A_b,\mu$ and $M_1$. We show that including
the corresponding phases strongly affects
the branching ratios of $\tilde t_{1,2}$ and $\tilde b_{1,2}$
decays in a large domain of the MSSM parameter space.
This could have an important impact on the search for $\st_{1,2}$ 
and $\sb_{1,2}$ and the determination of the underlying MSSM 
parameters at future colliders.
\end{abstract}

\newpage


\section{Introduction}

Many phenomenological studies on SUSY particle searches have been
performed in the Minimal Supersymmetric Standard Model (MSSM)
with real SUSY parameters. In general, however, some of the SUSY 
parameters may be complex, in particular the higgsino mass parameter 
$\mu$, the gaugino mass parameters $M_{1,2,3}$ and the trilinear scalar 
coupling parameters $A_f$ of the sfermions $\tilde{f}$. The SU(2) 
gaugino mass parameter $M_2$ can be chosen real after an appropriate 
redefinition of the fields. Not only the CP-violating observables 
(such as fermion EDMs) but also the CP-conserving observables 
(such as cross sections and decay branching ratios) depend on the
phases of the complex parameters, because in general the
mass-eigenvalues and the couplings of the SUSY particles involved are
functions of the underlying complex parameters. For example, the decay
branching ratios of the staus $\tilde{\tau}_{1,2}$ and $\tau$-sneutrino
$\tilde{\nu}_\tau$ can be quite sensitive to the complex phases of the stau
and gaugino-higgsino sectors \cite{CPslepton}. Therefore, in a complete 
phenomenological analysis of production and decays of the SUSY particles 
one has to take into account that $A_f$, $\mu$ and $M_{1,3}$ can be complex. 
In this article based on \cite{CPsquarkPLB,CPsquarkPRD} we study the 
effects of the phases of $A_t$, $A_b$, $\mu$ and $M_1$ on the decay 
branching ratios of the stops $\tilde{t}_{1,2}$ and sbottoms 
$\tilde{b}_{1,2}$ with $\tilde{q}_1$ ($\tilde{q}_2$) being the lighter 
(heavier) squark. We take into account the explicit CP violation 
in the Higgs sector. 

\section{SUSY CP Phase Dependences of Masses, Mixings and Couplings}

In the MSSM the squark sector is specified by the mass matrix in the basis 
$(\sq_L, \sq_R)$ with $\sq=\st$ or $\sb$
\begin{equation}
  {\cal M}^2_{\sq}= 
     \left( 
            \begin{array}{cc} 
                m_{\sq_L}^2 & a_q^* m_q \\
                a_q m_q     & m_{\sq_R}^2
            \end{array} 
     \right)       
                                                           \label{eq:a}
\end{equation}
with
\begin{eqnarray}
  m_{\sq_L}^2 &=& M_{\ti Q}^2 
                  + m_Z^2\cos 2\beta\,(I_3^{q_L} - e_q\sin^2\t_W) 
                  + m_q^2,                                 \label{eq:b} \\
  m_{\sq_R}^2 &=& M_{\{\ti U,\ti D\}}^2  
                  + m_Z^2 \cos 2\b\, e_q\, \sin^2\t_W + m_q^2, 
                                                           \label{eq:c} \\[2mm]
  a_q m_q     &=& \left\{ \begin{array}{l}
                     (A_t - \mu^*\cot\beta) \; m_t~~(\sq=\st)\\
                     (A_b - \mu^*\tan\beta) \; m_b~~(\sq=\sb)
                          \end{array} \right.              \label{eq:d} \\
              &=& \,\, |a_q m_q| \, e^{i\varphi_{\sq}} \,\,
                      (-\pi < \varphi_{\sq} \leq \pi).
                                                           \label{eq:e}
\end{eqnarray}
Here $I_3^q$ is the third component of the weak isospin and $e_q$ the 
electric charge of the quark $q$.
$M_{\ti Q,\ti U,\ti D}$ and $A_{t,b}$ are soft SUSY--breaking 
parameters and $\tan\b = v_2/v_1$ with $v_1$ $(v_2)$ being the vacuum 
expectation value of the Higgs field $H_1^0$ $(H_2^0)$. 
We take $A_q$ ($q=t,b$) and $\mu$ as complex parameters: 
$A_q = |A_q| \, e^{i\varphi_{A_q}}$ 
and $\mu = |\mu| \, e^{i\varphi_{\mu}}$ with 
$-\pi < \varphi_{A_q,\mu} \leq \pi $. 
Diagonalizing the matrix (\ref{eq:a}) one gets the mass eigenstates $\sq_1$ 
and $\sq_2$
\begin{equation}
     \left( \begin{array}{c} 
                \sq_1 \\
                \sq_2
            \end{array} \right) = R^{\sq} 
     \left( \begin{array}{c} 
                \sq_L \\
                \sq_R
            \end{array} \right) = 
     \left( \begin{array}{cc} 
                e^{i\varphi_{\sq}} \cos\t_{\sq} & \sin\t_{\sq} \\
                -\sin\t_{\sq}  & e^{-i\varphi_{\sq}} \cos\t_{\sq}
            \end{array} \right) 
     \left( \begin{array}{c} 
                \sq_L \\
                \sq_R
            \end{array} \right)
                                                         \label{eq:f}
\end{equation}
with the masses $m_{\sq_1}$ and $m_{\sq_2}$ ($m_{\sq_1} < m_{\sq_2}$), 
and the mixing angle $\t_{\sq}$
\begin{eqnarray}
  m_{\sq_{1,2}}^2 &=& \frac{1}{2}(m_{\sq_L}^2 + m_{\sq_R}^2 
    \mp \sqrt{ (m_{\sq_L}^2 - m_{\sq_R}^2)^2 + 4|a_q m_q|^2 }), 
                                                           \label{eq:g} \\
  \t_{\sq} &=& \tan^{-1}(|a_q m_q|/(m_{\sq_1}^2 - m_{\sq_R}^2)) \quad 
                                         (-\pi/2 \leq \t_{\sq} \leq 0).
                                                           \label{eq:h} 
\end{eqnarray}
The $\sq_L - \sq_R$ mixing is large if $|m_{\sq_L}^2 - m_{\sq_R}^2| 
\lsim |a_q m_q|$, which may be the case in the $\st$ sector due to the large 
$m_t$ and in the $\sb$ sector for large $\tan\b$ and $|\mu|$.

We assume that the gluino mass $m_{\sg}=M_3$ is real. 
We write the U(1) gaugino mass $M_1$ as 
$M_1 = |M_1| e^{i\varphi_1} \, (-\pi < \varphi_1 \leq \pi)$. 
Inspired by the gaugino mass unification we take 
$|M_1| = (5/3) \tan^2\t_W M_2$ and $m_{\sg} = (\a_s(m_{\sg})/\a_2) M_2$. 
In the MSSM Higgs sector with explicit CP violation the neutral Higgs mass 
eigenstates $H_1^0, H_2^0$ and $H_3^0$ $(m_{H_1^0} < m_{H_2^0} < m_{H_3^0})$ 
are mixtures of CP-even and CP-odd states. For the radiatively corrected masses 
and mixings of the Higgs bosons we use the formulae of Ref.\cite{CPhiggs}. 

Possible important decay modes of $\st_{1,2}$ are: 
\beqaa
  \st_1 & \to & t \sg \, , t \nt_i \, , \, b \chp_j \, , \, 
                \sb_1 W^+ \, , \, \sb_1 H^+
                                                           \label{eq:i} \\
  \st_2 & \to & t \sg \, , t \nt_i \, , \, b \chp_j \, , \,
                \st_1 Z^0 \, , \, \sb_{1,2} W^+ \, , \, \st_1 H_k^0 \, , \, 
                \sb_{1,2} H^+. 
                                                           \label{eq:j} 
\eeqaa
Those of $\sb_{1,2}$ are analogous. 
The CP phase dependences of the decay widths stems from those of the involved 
mass-eigenvalues, mixings and couplings among the interaction-eigenfields. 
In \cite{CPsquarkPLB,CPsquarkPRD} it is found that the masses, mixings and 
couplings of the involved SUSY particles (stops, sbottoms, charginos 
$\ch_i$ ($m_{\ch_1}<m_{\ch_2}$), neutralinos $\nt_j$ 
($m_{\nt_1}< ... < m_{\nt_4}$), and Higgs bosons) can be very sensitive to the 
CP phases in a large region of the MSSM parameter space. For the $\sq_i$  
($\sq = \st, \sb$) sectors we find:
\begin{enumerate}
      \item The mass-eigenvalues $m_{\sq_{1,2}}$ are sensitive to the phases 
            $(\varphi_{A_q}, \varphi_{\mu})$ 
            via $\cos(\varphi_{A_q} + \varphi_{\mu})$
            if and only if $|a_q m_q| \sim (m_{\sq_L}^2 + m_{\sq_R}^2)/2$ 
            {\em and} $|A_q| \sim |\mu| C_q$ (with $C_t = \cot\b$ and 
            $C_b = \tan\b$) (see Eqs. (\ref{eq:d}),(\ref{eq:g})).
      \item The $\sq$-mixing angle $\t_{\sq}$ (given by 
            $\tan 2\t_{\sq} = 2|a_q m_q|/(m_{\sq_L}^2 - m_{\sq_R}^2)$ ) 
            is sensitive to $(\varphi_{A_q}, \varphi_{\mu})$ via 
            $\cos(\varphi_{A_q} + \varphi_{\mu})$  if and only if 
            $2|a_q m_q| \gsim |m_{\sq_L}^2 - m_{\sq_R}^2|$ {\em and} 
            $|A_q| \sim |\mu| C_q$ (see Eqs. (\ref{eq:d}),(\ref{eq:g}),
            (\ref{eq:h})). 
      \item The $\sq$-mixing phase $\varphi_{\sq}$ in Eq.(\ref{eq:e}) is 
            sensitive to 
            ($\varphi_{A_q}$, $\varphi_{\mu}$), $\varphi_{A_q}$ and 
            $\varphi_{\mu}$ if $|A_q| \sim |\mu| C_q$, $|A_q| \gg |\mu| C_q$ 
            and $|A_q| \ll |\mu| C_q$, respectively. 
            For large squark mixing the term
            $\propto \sin 2\theta_{\tilde{q}} 
            \cos\varphi_{\tilde{q}}$ can result in a
            large phase dependence of the decay widths \cite{CPsquarkPRD} 
            (see Eq.(\ref{eq:f})).
\end{enumerate}
Therefore, we expect that the widths (and hence the branching ratios) of 
the $\st_{1,2}$ and $\sb_{1,2}$ decays are sensitive to the phases 
($\varphi_{A_{t,b}}$, $\varphi_{\mu}$, $\varphi_1$) in a large region of 
the MSSM parameter space. 

\section{Numerical Results}

In order to improve the convergence of the perturbative expansion 
\cite{improvedQCDcorr} we calculate the 
tree-level widths by using the corresponding tree-level couplings defined 
in terms of ``effective" MSSM running quark masses $m_{t,b}^{run}$ 
(i.e. those defined in terms of 
the effective running Yukawa couplings $h_{t,b}^{run}$ 
$\propto m_{t,b}^{run}$). For the kinematics, e.g., for 
the phase space factor we use the on-shell masses obtained by using the 
on-shell (pole) quark masses $M_{t,b}$. 
We take $M_t=175$ GeV, $M_b=5$ GeV, $m_t^{run}=150$ GeV, and $m_b^{run}=3$ GeV. 
We fix $|A_t|=|A_b| \equiv |A|$ and $M_2=300$ GeV, i.e. $m_{\sg}=820$ GeV. 
In our numerical study 
we take $\tan\b$, $m_{\st_1}$, $m_{\st_2}$, $m_{\sb_1}$, $|A|$, $|\mu|$, 
$\varphi_{A_t}$, $\varphi_{A_b}$, $\varphi_{\mu}$, $\varphi_1$ and 
$m_{H^+}$ as input parameters, where $m_{\st_{1,2}}$ and $m_{\sb_1}$ are 
the on-shell squark masses.  
Note that for a given set of the input parameters we 
obtain two solutions for ($M_{\ti Q}$, $M_{\ti U}$) corresponding to 
the two cases $m_{\st_L} \geq m_{\st_R}$ and $m_{\st_L} < m_{\st_R}$ 
from Eqs. (\ref{eq:a})-(\ref{eq:d}) and (\ref{eq:g}) with $m_t$ replaced 
by $M_t$. In the plots we impose the conditions as described in 
\cite{CPsquarkPLB,CPsquarkPRD} in order to 
respect experimental and theoretical constraints, such as the LEP limits 
and $b \to s \gamma$ data.

\noi
The experimental limits on the EDMs of electron, neutron and $^{199}$Hg 
strongly constrain the SUSY CP phases. Here we adopt the scenario of 
\cite{Nelson} where the first two generations of the sfermions are very 
heavy and hence $\varphi_{A_{t,b}}$, $\varphi_{\mu}$ and $\varphi_1$ are 
practically unconstrained. We have also checked that the electron and neutron 
EDM constraints at two-loop level \cite{Pilaftsis} are fulfilled in the 
numerical examples studied in this article.

In Fig.1 we show the $\varphi_{A_t}$ dependence of the $\st_2$ decay 
branching ratios for $\varphi_\mu=0$ and $\pi/2$ with $\tan\b$=8, 
($m_{\st_1}$,$m_{\st_2}$,$m_{\sb_1}$)=(400,700,200) GeV, $|A|=800$ GeV, 
$|\mu|=500$ GeV, $\varphi_{A_b}$=$\varphi_1$=0, and $m_{H^+}=600$ GeV 
in the case $m_{\st_L} \geq m_{\st_R}$. The case $m_{\st_L} < m_{\st_R}$ 
leads to similar results. We see that the $\st_2$ decay branching 
ratios are very sensitive to $\varphi_{A_t}$ and depend significantly on 
$\varphi_{\mu}$. 
For large $\tan\b (\gsim 15)$ we have obtained results similar to those 
for $\tan\b$=8 \cite{CPsquarkPLB,CPsquarkPRD}. 
\begin{figure}[t]
\begin{center}
\scalebox{0.5}[0.8]{\includegraphics{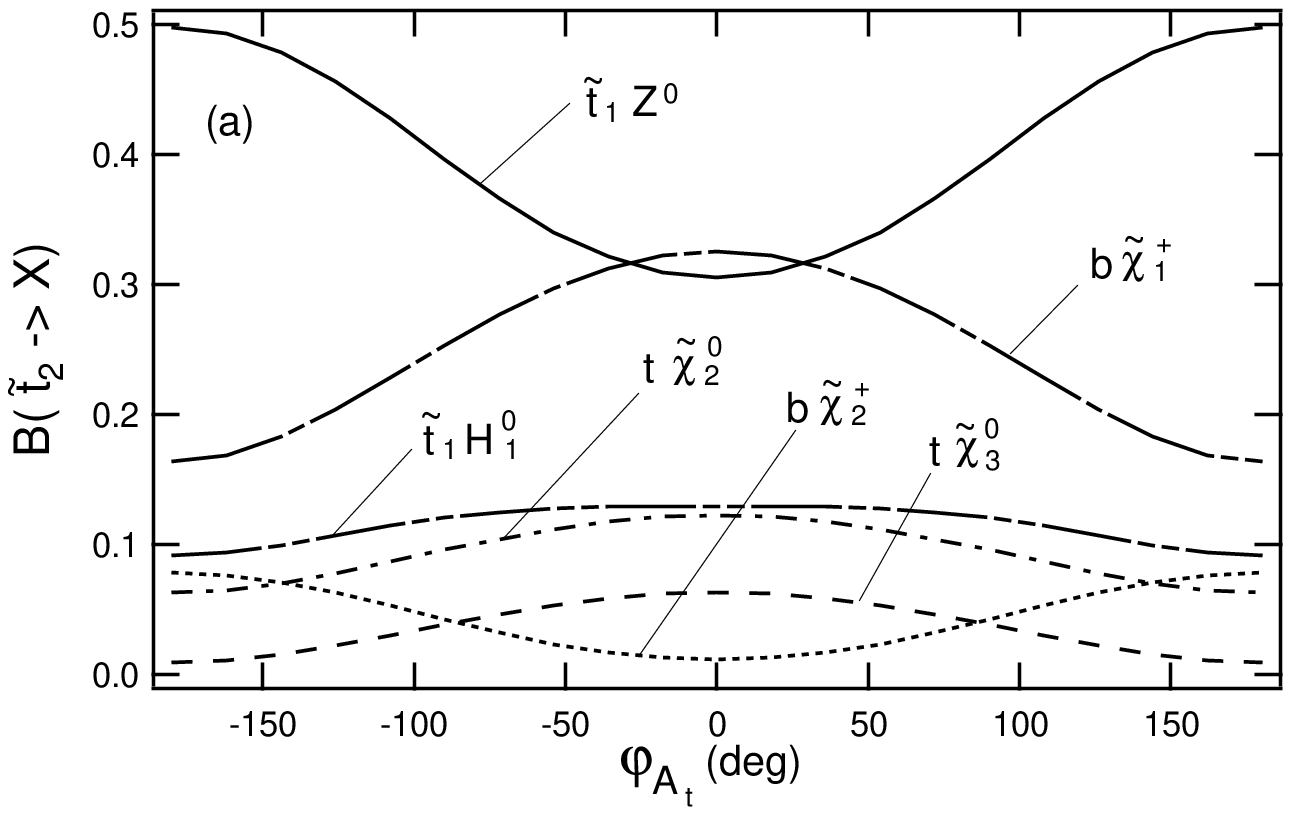}} 
\scalebox{0.5}[0.8]{\includegraphics{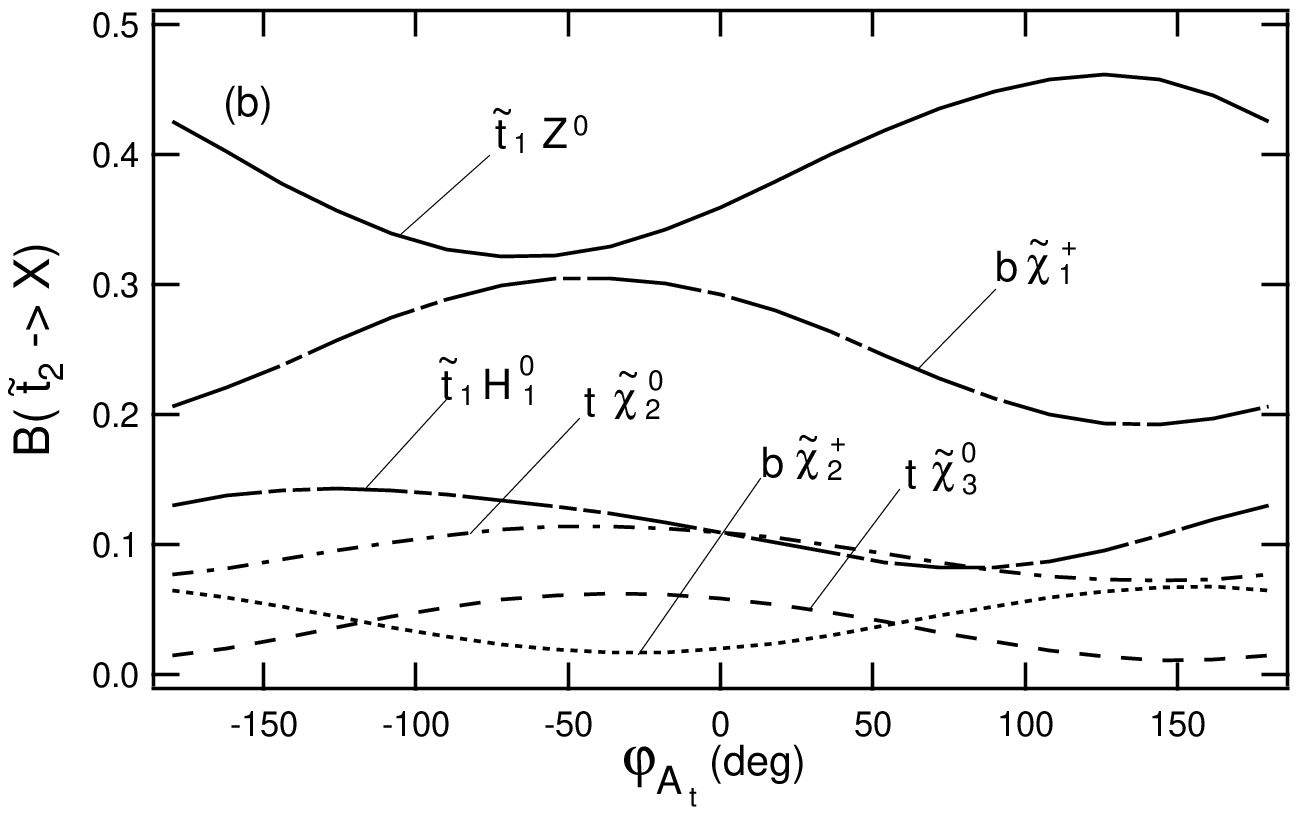}} 
\end{center}
\vspace{-5mm}
\caption{\label{Bst2phiAt} 
$\varphi_{A_t}$ dependence of the $\st_2$ decay 
branching ratios for $\varphi_\mu=0$ (a) and $\pi/2$ (b) with $\tan\b$=8, 
($m_{\st_1}$,$m_{\st_2}$,$m_{\sb_1}$)=(400,700,200) GeV, $|A|=800$ GeV, 
$|\mu|=500$ GeV, $\varphi_{A_b}$=$\varphi_1$=0, and $m_{H^+}=600$ GeV 
in the case $m_{\st_L} \geq m_{\st_R}$. Note that the $\st_1 H_{2,3}^0$ 
and $\sb_1 H^+$ modes are kinematically forbidden here.}
\end{figure}
For $\st_1$ and  $\sb_{1,2}$ decays we have obtained results similar to those 
for the $\st_2$ decays \cite{CPsquarkPLB,CPsquarkPRD}. 
We have also found that for small $\tan\b (\lsim 8)$ the $\st_1$ decay can 
be fairly sensitive to $\varphi_1$ \cite{CPsquarkPLB}.

\section{Parameter Determination}
 
We now study to which extent one can extract
the underlying MSSM parameters (such as $\tan\b$, $A_{t,b}$ etc.) 
from measured observables (such as masses, cross sections and branching 
ratios). The observables are functions of the MSSM parameters. Therefore, 
in case the number of measured observables is larger than that of the MSSM 
parameters, the whole data set of the observables over-constrains the MSSM 
parameters. In this case the MSSM parameters and their errors can be 
determined by a $\chi^2$ fit to the experimental data of the observables. 
In the fit the errors of the observables are propagated to those of the 
MSSM parameters. The assumptions for the expected experimental errors of 
the observables measured at TESLA, CLIC and LHC are described in 
\cite{CPsquarkPRD}. Our strategy for the parameter determination 
is as follows:
\begin{enumerate}
\item Take a specific set of values of the underlying MSSM
  parameters. 
\item Calculate the masses of
  $\tilde t_i$, $\tilde b_i$, $\tilde \chi^0_j$, $\tilde \chi^\pm_k$,
  $H_{\ell}^0$, the production cross sections for $e^+ e^- \to \tilde t_i
  \bar{\tilde t}_j$, and $e^+ e^- \to \tilde b_i \bar{\tilde b}_j$, and 
  the branching ratios of the $\tilde t_i$ and $\tilde b_i$ decays, 
  and estimate the expected experimental errors of these observables. 
\item Regard these calculated values as real experimental data with definite
  errors.
\item Determine the underlying MSSM parameters and their errors from the
  ``experimental data'' by a $\chi^2$ fit. 
  \end{enumerate}

We consider two scenarios, 
one with small $\tan\beta$ and one with large $\tan\beta$. 
The small $\tan\beta$ scenario is characterized by the following underlying 
MSSM parameters: 
$M_{\ti D} =$   169.6~GeV, $M_{\ti U} =$   408.8~GeV, 
$M_{\ti Q} =$   623.0~GeV,
$|A_t| = |A_b| = $ 800 GeV, 
$\varphi_{A_t} = \varphi_{A_b} = \pi/4$, $\varphi_1=0$,
$M_2 =$  300~GeV, $\mu = -350$~GeV, $\tan \beta = 6$, 
 $m_{\tilde g} =$ 1000~GeV, and $m_{H^\pm} =$  900~GeV.
(Here we do not assume the unification relation between
$m_{\tilde g}$ and $M_2$.)
The large $\tan\beta$ scenario is specified by: 
 $M_{\ti D} =$   360.0~GeV,
 $M_{\ti U} =$   198.2~GeV,
 $M_{\ti Q} =$   691.9~GeV,
$|A_t| = $ 600 GeV, 
$\varphi_{A_t} = \pi/4$,
$|A_b| = $ 1000 GeV, 
$\varphi_{A_b} = 3 \pi/2$,
$\varphi_1 = 0$,
 $M_2 =$  200~GeV,
$\mu = -350$~GeV,
  $\tan \beta =$   30,
 $m_{\tilde g} =$ 1000~GeV, and
  $m_{H^\pm} =$  350~GeV.
The resulting values of the observables and their expected experimental 
errors are shown for the two scenarios in \cite{CPsquarkPRD}. We regard 
these calculated values as real experimental data with definite errors. 
We determine the underlying MSSM parameters and their 
errors from the ``experimental data'' on these observables by 
a $\chi^2$ fit. The results obtained are shown in Table~2 of 
\cite{CPsquarkPRD}. As one can see, all parameters except $A_b$ can be 
determined rather precisely.
$\tan\beta$ can be determined with an error of about 3\% in both scenarios. 
The relative error of the squark mass parameters squared 
$M_{\ti Q,\ti U,\ti D}^2$ is in the range of 1\% to 2\%. 
$Re(A_t)$ and $Im(A_t)$  can be measured within an error of 
2 -- 3\% independently of $\tan\beta$. 
The situation for $A_b$ is considerably worse: in case of small $\tan\beta$
one gets only an order of magnitude estimate. The reason is that
both the bottom squark mixing angle and the bottom squark couplings depend
only weakly on  $A_b$ for small $\tan\beta$. In case of large 
$\tan\beta$ the situation improves somewhat in particular for
the imaginary part of $A_b$. 
A $\chi^2$ fit using only real MSSM parameters would result in a totally 
wrong parameter determination with a significantly larger value of $\chi^2$. 
We have found that the analogous fit procedure using only 
real MSSM parameters gives a much larger value for $\chi^2$:
$\Delta\chi^2=286.6$ with DOF=61 for the scenario with $\tan\beta=6$ and 
$\Delta\chi^2=22.5$ with DOF=61 for the scenario with $\tan\beta=30$, 
where DOF is 'degree of freedom'.

\section{Conclusion}

We have studied the decays of $\tilde t_{1,2}$ and $\tilde b_{1,2}$ in the 
MSSM with complex parameters $A_{t,b},\mu$ and $M_1$. We have shown that 
including the CP phases strongly affects the branching ratios of 
$\tilde t_{1,2}$ and $\tilde b_{1,2}$ decays in a large domain of the MSSM 
parameter space. This could have an important impact on the search for 
$\st_{1,2}$ and $\sb_{1,2}$ and the determination of the underlying 
MSSM parameters at future colliders. 


%
\end{document}